\documentclass{epl}
\usepackage{latexsym}

\title{The failure of determinism and the limits of of Bohm hidden variables theory}

\shorttitle{The failure of determinism}

\author{R. Lapiedra\thanks{E-mail address: ramon.lapiedra@uv.es (R. Lapiedra). Telephone: 96 3543077}}

\institute{Department of Astronomy and Astrophysics, University of
Valencia, 46100 Burjassot (Valencia), Spain}

\pacs{03.65.Ud}{Entanglement and quantum non-locality}
\pacs{03.65.Ta}{Foundations of Quantum Mechanics, measurement
theory}

\begin{document}

\maketitle

\begin{abstract}
We give a counter example to show that determinism as such is in
contradiction to quantum mechanics. More precisely, we consider a
simple quantum system and its environment, including the
measurement device, and make the assumption that the time
evolution of the global system, the quantum system plus this
environment, is deterministic, i. e., its time evolution is given by
a dynamical trajectory from some initial conditions. From this, we
prove a type of Bell inequalities which are violated by quantum
mechanics, reaching the conclusion that our quantum system evolves
in a non deterministic way. In order to seize the interest of this
conclusion, one must realize that it cannot be reached from the
present experimental violation of Bell inequalities predicted by
quantum mechanics, since this violation is compatible with
non-local realism.
\end{abstract}

\section{1. Introduction}\label{intro}
Let us consider the time evolution of an isolated
physical system, either classical or quantum. Does it always
exist, as a matter of principle, a trajectory which, from some
initial conditions, gives the values of the different quantities
of the system at any time, as it is, for example, the case in Newtonian mechanics? This is the problem of determinism as
such, i. e., determinism without any supplementary conditions, in
the natural world. Obviously, determinism as such, or simply
determinism for short, is sometimes present in the world, but the
problem is if determinism is always present irrespective of our
capacity of prediction in practice. Thus, for our purpose here it
will be enough to give an example where determinism fails, meaning
that it enters in contradiction to quantum predictions. This is
what we are going to do along the present paper.

Nevertheless, before this, we will consider two objections that,
from the very beginning, could be raised against this goal.

The first one would be to say that we are just referring to the
time evolution of an isolated physical system, while one could
always doubt if there is any physical system which can be
considered really isolated.

We will see in the following Section that this objection can be
easily neutralized, so let us consider in some detail the second
objection, which could be phrased as follows: is not the above
claimed fail of determinism denied by the kind or realism still
allowed by the reported experimental violation \cite{Aspect}
\cite{Rowe} of Bell inequalities \cite{Bell} \cite{Clauser}?
Notice that, as it is very well known, this violation entails the
fail of local realism, but not the fail of the non-local one, i.
e., the violation does not deny realism as such. In other words,
we could keep realism as such in front of this experimental
violation if we renounced to keep locality by leaving aside
special relativity, which, on the other side, would raise
tremendous experimental and theoretical problems. But, we do not
need to have recourse to such a strong assumption in order to
leave aside locality since, as pointed out by Bell \cite{Bell2},
we also could rely on a vast \emph{conspiracy} which, in the frame
of determinism, would arrange the causally unconnected hidden
variable values in order to produce the correlations, between the
two particle measurement outcomes, which are responsible for the
observed violation of Bell inequalities. In other words, the
\emph{conspiracy} would make compatible realism as such with the
experimental violation of Bell's inequalities. Then, could not
this \emph{conspiracy} also be invoked to see, from the very
beginning, that the above claimed fail of determinism as such must
be erroneous? We will see in the next Section that this is not the
case by actually proving, in a certain case, that determinism
enters in contradiction to quantum mechanics. The reason why non
local realism can be reconciled with the violation of Bell
inequalities but not with determinism is that the determinism
assumption is a stronger condition than realism assumption in the
case of an EPR experiment. In the last case, one assumes the
existence of some hidden variables values behind the outcomes of a
couple of measurements performed on two entangled particles,
without asking that these hidden values remain the same after
performing this couple of measurements. On the contrary, in the
case of determinism, one assumes some \emph{common} hidden variable
values (the unknown initial conditions) behind \emph{all} the successive
spontaneous responses of the isolated system along a lapse of
time.

Perhaps some reader will find nonsense to resort to such a
fantastic idea as the one of the above \emph{conspiracy} to keep
non-locality. Nevertheless, the author does not know what could be
more fantastic, wether this \emph{conspiracy} or what is at stake
in the present paper, i. e., the absence of determinism. Thus,
from this point of view, to consider the possibility of this
\emph{conspiracy}, as it is made in the present paper, is not only
convenient but even necessary.

Finally, one could also ask whether this claimed fail of
determinism could encompass, for example, the case of a quantum
particle, since it is widely claimed that Bohm \cite{Bohm} has
actually built a non-local realistic theory which reproduces all
quantum predictions: a non-local realistic theory where any
quantum particle would have its own well defined trajectory, at
least in the absence of measurement, and so a deterministic theory
to a certain extent. Then, perhaps this determinism could still be
preserved in Bohm's theory framework, when successive measurements
on the particle are present. However, in Section 3, we will see
that, whatever could be the merits of the Bohm theory, the theory
cannot produce this determinism in the presence of measurement and
at the same time agree with quantum mechanics (QM), since this
determinism enters in contradiction to QM.

\section{2. Determinism as such, the entailed Bell inequalities and its violation}\label{sec:2}
Let it be a free half one spin particle, on which we perform the
following ideal experiment: we first fix three space directions
given by the unit 3-vectors $\vec{a}$, $\vec{b}$ and $\vec{c}$. Then, we
measure successively the particle spin, each time along one of the
precedent directions randomly selected.

Let us consider the physical system composed of the half one spin
particle plus the affecting environment including the experimental device with the part of this
device that selects randomly the measurement direction among the
three initially fixed directions. Assume that this global system remains isolated. Now, as it happens for example with the Newtonian determinism (where the initial position and velocity, i. e., the initial conditions, allow us to know the entire particle trajectory), assume strict determinism
in the evolution of this global system. By this I mean that not
only the successive spin measurement outcomes, say $\pm{1}$, are determined
from the initial conditions, but that the successively
selected measurement directions are determined too, let this random selection be
performed by a \emph{free} subject or by any other means. (Notice
that I do not put any restrictions on these assumed initial
conditions. In particular, they could range over space-time
regions not causally connected).

Let us be more precise. Let us denote as $\lambda$ those initial conditions. Imagine that we have a set of such systems, differing among them by different initial conditions, i. e., by different $\lambda$ values. We will perform two consecutive measurements on every prepared system of the above set (each system labeled by a value of $\lambda$). These two consecutive spin measurements will be performed, respectively, at two times randomly selected among three standing times $t_1, t_2$, and $t_3$, along two directions randomly selected among three standing directions given by the above three unit 3-vectors $\vec{a}$, $\vec{b}$, $\vec{c}$. Denote by $S$ the values, $\pm 1$, of the measurement outcomes. Because of the determinism postulate, there exist some initial conditions, i. e., some parameter values, $\lambda$, such that we can write $S=S(\lambda, t_i, \vec{x}(t_i))$, $i=1,2,3$, for these outcomes, corresponding to a given value of $\lambda$ and $t_i$, with $\vec{x}(t_i)\in (\vec{a}, \vec{b}, \vec{c})$. Notice that the notation is redundant, since because of the postulate of determinism once we have fixed $\lambda$ and $t_i$ the value of $S$ becomes determined. Notice too that, because of the postulate of determinism, shifting as an all the three above times while keeping $\lambda$ fixed is equivalent to keep the three original times while changing in a convenient way the original initial conditions.

Now, we are going to mimic the original proof of Bell inequalities \cite{Bell} in order to prove similar inequalities for our measurement outcomes. Let us consider the following three expectation values

\begin{equation}
P(a,b)=\int d\lambda \rho (\lambda) S(\lambda, t_1, \vec{a})S(\lambda, t_2, \vec{b}), \label{expectvaluab}
\end{equation}

\begin{equation}
P(a,c)=\int d\lambda \rho (\lambda) S(\lambda, t_1, \vec{a})S(\lambda, t_3, \vec{c}), \label{expectvaluac}
\end{equation}

\begin{equation}
P(b,c)=\int d\lambda \rho (\lambda) S(\lambda, t_2, \vec{b})S(\lambda, t_3, \vec{c}), \label{expectvalubc}
\end{equation}
where $\rho(\lambda)$ stands for the probability density of the $\lambda$ values.

But, we could doubt if we can put, as we have done, the same output value, $S(\lambda, t_2, \vec{b})$ in (\ref{expectvaluab}) and in (\ref{expectvalubc}), since in the first case this value is paired up with $S(\lambda, t_1, \vec{a})$ whereas in the second case is paired up with the different $S$ value, $S(\lambda, t_3, \vec{c})$ (we could raise a similar question about the other two outcome values $S(\lambda, t_1, \vec{a})$ and $S(\lambda, t_3, \vec{c})$). Nevertheless, as we have just remarked, our postulate of determinism insures that, once $\lambda$ and $t_i$ have been given, the corresponding outcome $S$ is fixed irrespective of the other outcome to which is paired up. This allows us to put the same value $S(\lambda, t_2,\vec{b})$ in Eqs. (\ref{expectvaluab}) and (\ref{expectvalubc}).

Obviously, for a given time, $t_i$, we can have different measurement directions, i. e., different values, $\vec{a}$, $\vec{b}$, or $\vec{c}$, for the vector function $\vec{x}$, as far as we change $\lambda$ in a convenient way. This means that the $\lambda$ values which are present in (\ref{expectvaluab}), (\ref{expectvaluac}), (\ref{expectvalubc}), are the ones that assign the vectors $\vec{a}$, $\vec{b}$, $\vec{c}$, respectively, to the times $t_1, t_2, t_3$.

Then let us consider the difference

\begin{equation}
P(a,b)-P(a,c)=\int d\lambda \rho (\lambda) S(\lambda, t_1, \vec{a})[S(\lambda, t_2, \vec{b})-S(\lambda, t_3, \vec{c}). \label{difference}]
\end{equation}

Henceforth, the proof of our inequalities goes along the same lines as the proof of the original Bell inequalities in the Bell's seminal paper \cite{Bell}. First, since $S^2(\lambda, t_2,\vec{b})=1$, the above difference can be written as

\begin{equation}
P(a,b)-P(a,c)=\int d\lambda \rho (\lambda) S(\lambda, t_1, \vec{a})S(\lambda, t_2, \vec{b})[1-S(\lambda, t_2, \vec{b})S(\lambda, t_3, \vec{c}). \label{difference2}]
\end{equation}

Then, taking absolute values, we are led to the inequality

\begin{equation}
|P(a,b)-P(a,c)| \le \int d\lambda \rho (\lambda) [1-S(\lambda, t_2, \vec{b})S(\lambda, t_3, \vec{c}). \label{inequality}]
\end{equation}
that is, to the well known Bell inequality

\begin{equation}
|P(a,b)-P(a,c)| \le 1-P(b,c). \label{inequality2}
\end{equation}

Notice that, as it has been said in the Introduction, the postulate od determinism, which is behind the above proof, would loose all its likeliness for a non isolated system. Nevertheless,
if a new part, not considered up to here, of the general
environment were finally supposed to affect, sensibly enough, the
original system, the inequality would then refer to the new system
enlarged with this new environment part. Of course, it is to be
expected that the larger this affecting environment become, the
lesser become the possibility, if any, of violating the resulting
inequality.

But, coming back to the question of testing inequality
(\ref{inequality2}), whatever could be the difficulties
to perform the kind of experiment we are considering, the three
mean values in (\ref{inequality2}) can be calculated as the
corresponding expected values dictated by quantum mechanics. These
values become $P(a,b)=\vec{a}.\vec{b}$ \cite{Brukner}\cite{Lapiedra}
and similarly for $P(b,c)$and $P(c,a)$. Thus, inequality (\ref{inequality2})
becomes
\begin{equation}
|\vec{a}.\vec{b}-\vec{a}.\vec{c}|+\vec{b}.\vec{c}\le 1,
\label{inequality3}
\end{equation}
which is violated for $\vec{b}.\vec{c}=0$ and
$\vec{a}=(\vec{a}+\vec{c})/\sqrt{2}$, in which case the left hand
side of inequality (\ref{inequality3}) reaches the value
$\sqrt{2} $ .

Thus, for the global system consisting of a 1/2-spin particle plus its affecting environment, including
the measurement device selecting the
measurement directions, the assumed determinism enters in
contradiction with quantum mechanics.

This fail of determinism could come from the quantum system affected by some generic and hypothetical environment, or from the measurement device (or from both). Nevertheless, we
always could perform the direction selection by a well determined
procedure: one compatible with the equality of the three
probabilities of obtaining one or another direction from the three
initially fixed ones. Of course, this deterministic procedure is
not expected to change the measurement outcomes. As a result,
whatever could be the extent of a hypothetical environment acting
on the particle which is submitted to successive measurements, we
have proved that determinism for this particle and quantum
mechanics are incompatible. In other words, there is no trajectory
for this quantum particle whatever it were the kind of realism
associated to the initial conditions: local or non-local
realism, or realism of a \emph{conspiratory} nature.

Notice that inequality (\ref{inequality2}) could be applied to any system, macroscopic or not, with a random dichotomic response to three kinds of consecutive measurements, only one of these three measurements being randomly selected each time, provide that we assume determinism for the time evolution of the system. Then, looking for the possible violation of such inequality could lead to test determinism in the natural world, beyond the particular case of a half one spin particle we have just tested.

At this point, it would be valuable to compare our results with other similar well known results in the literature on the subject:

First, with the Kochen-Specker theorem \cite{Kochen}. According to the readable and simplified version of this theorem given in \cite{Cabello}, its two basic assumptions are \emph{counterfactual definiteness} ("two non-compatible observables can simultaneously have predefined values") and \emph{non-contextuality} ("the value of an observable ... does not depend on which other observables  (compatible with it) have simultaneously defined values"). Obviously, in our case the postulate of determinism implies counterfactual definiteness, but now non-contextuality does not make sense since the measured observables (successive spin measurements) are incompatible observables, aside the fact that its measurement outcomes are not independent among them, but strongly correlated by hypothesis. Then, our results and the Kochen-Specker theorem deal with different situations.

Next we consider "The free will theorem" \cite{Conway}. Here, the problem considered departs from the very beginning from the one considered in the present paper. While, in the frame of this theorem, the choice of directions in which to perform spin experiments is not a function of the information accessible to the experimenters, in our case this choice is assumed to be given in a deterministic way from some initial conditions, which could be known in principle by the experimenter, or even produced by his \emph{free} will provided that the different directions appear with the same probability.

Finally, in \cite{Leggett}, under the following three postulates, \emph{macroscopic realism per se}, \emph{noninvasive measurability} and \emph{induction}, Leggett claims to have proved some CHSH inequalities \cite{Clauser}, for the successive outcomes of a system with random dichotomic responses against four types of measurements. Our determinism postulate entails macroscopic realism per se, and noninvasive measurability, for the global system (the spin 1/2 particle plus the environment including the measurement device), but we have not needed to invoke the induction postulate to prove our Bell inequalities. Invoking this postulate in our case would entail restrictive conditions on the initial conditions, $\lambda$, that we have not needed to impose to prove our inequality (\ref{inequality2}).

Now, in according to what I have said at the end of the
Introduction, it seems that this claimed non existence of a
trajectory should also be considered from the point of view of the
Bohm hidden variable theory, in order to see that there is no
contradiction between the present claimed non determinism and this theory.
In the next section we explain why this is actually the case.

\section{3. The limits of Bhom theory of hidden variables}\label{sec:3}

Let us consider in detail to what extent is it really true that
Bohm's hidden variable theories (HVT) can predict the existence of
dynamical trajectories and at the same time be consistent with QM.
It is true that Bohm \cite{Bohm} proves that his theory gives the
same probability of finding a particle in a given position that QM
does. From this, he concludes that his "interpretation is capable
of leading in all possible experiments to identical predictions to
those obtained from the usual interpretation", that is to say, to those obtained from QM. Then, when considering an entangled
extended system, as in Einstein-Podolsky-Rosen experiments
(similar to the ones considered by Bell in his seminal papers),
Bohm assumes that his realism is non-local. In this way, his
non-local HVT can explain the observed violation of the ordinary
Bell inequalities, in agreement with QM, without having to give up
realism (see \cite{Bell} for example).

But is it always this way? Is it true that we can devise actual
non-local HVT that lead to the same predictions that QM, \emph{
for all conceivable experiments}? Let us make some considerations
in order to show why, from the very beginning, it is dubious that
HVT, even if allowing for non-local realism, could always agree
with QM to the extent of assigning a dynamical trajectory to each
quantum particle when measurements are present. That is, to the
extent of assuming determinism.

First of all, in these theories, each time that one performs a
measurement on the particle position, if one wants to complete,
beyond the obtained outcome, the precedent particle trajectory
with a new trajectory piece, one must provide the probability
density of the particle position just after this outcome. The
provided probability becomes the new initial probability. Then,
this new initial probability must be taken the same as the one
dictated by standard QM if we want the HVT to agree henceforth
with QM. After this, in the HVT framework, one does not need to
worry about how this initial probability evolves in time until one
performs a subsequent measurement, since, in the absence of any
measurement, HVT are just designed to predict the same probability
evolution as the one predicted by Schr\"odinger equation. But, as
we will see in a moment, the real point is that when some
consecutive different measurements are performed on the same
particle \cite{Lapiedra}, one expects to find some well definite
correlations among the corresponding outcomes: the correlations
dictated by QM that lead, for example, to the violation of
inequalities (\ref{inequality2}). As it is asserted without proof
(and without any preciseness about what type of realism would be
ruled out), in\cite{Streater}: "In QM, positions at different times
do not commute, so ... some correlations referring to positions at
different times fail to satisfy Bell's inequalities". As it has
been shown above, what is actually violated, because of the QM, is
some Bell-type inequality coming from the assumption of determinism
based in \emph{local or non-local realism}.

More precisely: either in QM or in HVT, the probabilities of each
measurement outcome is given by the corresponding initial quantum
state of the particle, just the state previous to the measurement.
In HVT, these initial quantum states are supplemented with the
assumed initial values of some non-local hidden variables,
$\lambda$, leading to the deterministic time evolution of the
system, in the absence of measurement. This determinism preserves,
as it must be, the quantum evolution of the outcome probabilities.
These initial $\lambda$ values can always be established and this
is the great triumph of Bohm theory. Nevertheless, the point here
is that these $\lambda$ values, which mimics so perfectly well the
quantum evolution of the above probabilities, in the absence of
measurement, have nothing to do with the explanation of the
quantum correlations which are behind the reported quantum
violation of inequalities (\ref{inequality2}). It has nothing to
do \emph{since these correlations have only to do with the quantum
fact that these initial $\lambda$ values need to be different
before and after a given measurement on the particle}, while they
have to be the same if one assumes uncritically that we always can
have deterministic dynamical trajectories for quantum particles in
the framework of HVT theories. In other words, from the very
beginning, and as we have done in the present paper, one should
ask wether those quantum correlations will always be compatible
with determinism, that is, with the assumption that some common
initial conditions, $\lambda$, local or not, are behind all the
successive measurements along the trajectory of a quantum
particle. Then, the answer to this question has to be negative
since, in the precedent Section, we have seen in detail how the
inequality (\ref{inequality2}), which comes from the assumption of
determinism, is violated by the corresponding quantum predictions.

Thus, if we mean by "trajectory" something more than a mere uninterrupted path,
even a zigzagging one, to require the existence of determining initial conditions,
and we accept QM, it seems that there is no room left "for models
that force Nature to mimic the concept of trajectory" as it is still
expected in \cite{Suarez}.

To summarize: according to the above discussions, either Quantum
Mechanics, or determinism as such, must be false. So, if on the
ground of its general success we accept QM, we must conclude that
determinism as such could contradict experiments, an statement
that would deserve being considered. Then the answer to the Leggett
question \cite{Leggett} whether "it is indeed realism rather than
locality which has to be sacrificed?" would be `yes'. All in all:
against Einstein's old dream, it seems that QM cannot be completed
to the extent to allow for the existence of trajectories for
quantum particles even when accounting for all its affecting
environment.

\begin{acknowledgements}
This work has been supported by the Spanish Ministerio de
Educaci\'on y Ciencia, MEC-FEDER project No.FIS2006-06062.
I also thank Dr Michael Hall by having made a number of useful
critical comments to the manuscript, and Prof. Eliseu Borr\`as by his reading and comments.
\end{acknowledgements}

\end{document}